\documentclass[useAMS,usenatbib]{mn2e}
\usepackage{amsmath}
\usepackage{amssymb}

	\usepackage{pslatex}
	\usepackage{amsmath}
	\usepackage{hyperref}
\usepackage{epsfig}
\usepackage{graphicx}
\usepackage{graphics}

\title[Influence of Photoelectrons]{Influence of Photoelectrons on the Structure and Dynamics
of the Upper Atmosphere of a Hot Jupiter}
\author[D. E. Ionov et al.]{D. E. Ionov\footnote{pereversi@gmail.com, @pantarktik in Telegram}, V. I. Shematovich, Ya. N. Pavlyuchenkov
\\
Institute of Astronomy, Russian Academy of Sciences, Moscow, Russia
}
\begin{document}
\maketitle

\begin{abstract}
A self-consistent, aeronomic model of the upper atmosphere of a ``hot
Jupiter'' including reactions involving suprathermal photoelectrons is
presented. This model is used to compute the height profiles of the gas
density, velocity, and temperature in the atmosphere of the exoplanet HD~209458b.
It is shown that including suprathermal electrons when computing
the heating and cooling functions reduces the mass loss rate of the
atmosphere by a factor of five.
\end{abstract}


\def\gtrsim{\mathrel{\hbox{\rlap{\hbox{\lower5pt\hbox{$\sim$}}}\hbox{$>$}}}}

\section{Introduction}

Studies of exoplanets have recently become one of the most active areas
of astrophysical research. Thousands of exoplanets have already been
discovered, and this number is continuously
growing~\cite{Schneider-2016}. Modern techniques make it possible not
only to detect an exoplanet and determine its orbital characteristics,
but also to obtain information about the parameters of its
atmosphere~\cite{Seager-2010}. Observations of transits and secondary
eclipses, as well as direct observations of exoplanets, can be used to
obtain spectra of upper atmospheric
layers~\cite{Vidal-Madjar-2003,Vidal-Madjar-2004,Linsky-2010}, making it
possible to determine their composition, structure, and dynamics.

The first exoplanets for which such observations were carried out were
``hot Jupiters'' --- exoplanets with masses comparable to that of Jupiter
located no more than 0.1 AU from their stars~\cite{Murray-Clay-2009}.
Spectral observations of the hot Jupiter HD 209458b --- the first
transiting exoplanet, as far as we are aware ---  obtained in 2003 showed
that this object is surrounded by an extended envelope of neutral
hydrogen extending beyond its Roche lobe~\cite{Vidal-Madjar-2003}.
Similar envelopes were later discovered for other hot Jupiters, such as
HD 189733b~\cite{Lecavelier-2010}, and WASP-12b~\cite{Fossati-2010a, 
Fossati-2010b}, and also for the ``warm Neptune'' GJ~436b~\cite{Ehrenreich-2015}.
It was established that the atmospheres of
planets with such envelopes should undergo a gas-dynamical outflow of
matter. The mass loss rate for the exoplanet HD 209458b was estimated to
be $10^{10}$  g/s~\cite{Vidal-Madjar-2003, Lammer-2003,Lammer-2009}.

The outflow of the atmosphere of HD~209458b has been studied by various
authors using gas-dynamical simulations~\cite{Yelle-2004,Munoz-2007,
Koskinen-2013,Shaikhislamov-2014} taking into account a variety of
chemical reactions. In all these studies, a one dimensional system of
gas-dynamical equations was solved; the main differences between the
various published models concern the composition of the atmosphere and
the boundary conditions applied. The atmosphere was taken to be composed
of atomic hydrogen, molecular hydrogen, and helium in
\cite{Yelle-2004}. The presence of minor species was taken
into account in~\cite{Koskinen-2013} and
\cite{Munoz-2007}. \cite{Koskinen-2013} included
 C, C$^+$, O, O$^+$, N, N$^+$, Si$^+$, Si, and Si$^{2+}$,
in their model, but assumed an absence of molecular hydrogen in
the upper atmosphere, while~\cite{Munoz-2007} also
included molecules comprised of C, O, N, and D
atoms.

A free-outflow condition is usually adopted as an external boundary
condition, when the values of the gas-dynamical parameters at the outer
boundary are extrapolated from adjacent cells. The external pressure was
specified at the upper boundary in ~\cite{Munoz-2007}. Fixed boundary
conditions are usually imposed at the lower boundary~\cite{Yelle-2004,
Munoz-2007}, although an inflow condition was imposed
in~\cite{Koskinen-2013}. The results obtained in these various studies
show that the upper atmosphere can be heated to temperatures exceeding
10000~K, although the equilibrium temperature for the planet is 1300~K.
These models can produce outflow rates corresponding to observational
estimates. However, models based exclusively on the gas-dynamical
equations are insufficiently accurate to describe heating processes in
the upper atmosphere, where the velocity distribution of the gas
particles is non- Maxwellian~\cite{Shematovich-2015}.

The main reason for the gas-dynamical outflow of the atmosphere of a hot
Jupiter is heating by the star’s radiation~\cite{Lammer-2003,
Lammer-2009,Shaikhislamov-2014,Tian-2005} at 1--100 nm (so-called XUV
radiation). This wavelength interval is conventionally divided into the
extreme ultraviolet (EUV, 10--100 nm) and soft X-ray (1--10 nm) ranges.
XUV radiation is absorbed during atomic-hydrogen and helium ionization
reactions, and also during the ionization, dissociation, and dissociative
ionization of molecular hydrogen~\cite{Shematovich-2010,Shematovich-2014,
Ionov-2015}. These processes are described by the reaction equations
\begin{equation}
\begin{array}{l}
 H_2 + h\nu (e_p) \rightarrow H_2^+ + e + (e_p) \\
 H_2 + h\nu (e_p) \rightarrow H(1s) + H(1s, 2s, 2p) \\
 H_2 + h\nu (e_p) \rightarrow H(1s,2p) + H^+ + e + (e_p) \\
 H + h\nu (e_p) \rightarrow H^+ + e + (e_p) \\
 He + h\nu (e_p) \rightarrow He^+ + e + (e_p) 
\end{array}
\label{eq_ioniz}
\end{equation}
where $h\nu$  represents an XUV photon,  $e_p$ a photoelectron,
and $e$ a secondary electron.

Some of the energy of an absorbed photon, equal to or exceeding the
ionization or dissociation energy, goes into the internal energy of the
matter, while the remainder is transformed into the kinetic energy of the
reaction products, to a large degree the kinetic energy of the electrons.
If the energy of a photoelectron that is produced is sufficiently high
--- more than an order of magnitude above the thermal energy (it is a
so-called suprathermal electron), it can participate in secondary
reactions leading to the ionization and excitation of components in the
atmosphere. All of the electrons’ initial kinetic energy is spent in this
case. Another channel for the loss of the initial energy of
photoelectrons is elastic collisions, which transform this energy into
heat. Thus, part of the energy of the photoelectrons goes into internal
energy, and part into heating the atmosphere. The effect of reactions
involving suprathermal photoelectrons is to decrease the fraction of the
energy of the star’s radiation that goes into heating the gas.

The role of photoelectrons in the gas heating is taken into account
in~\cite{Yelle-2004,Munoz-2007, Koskinen-2013,Shaikhislamov-2014} using
an adjustable coefficient called the heating efficiency. Physically, the
heating efficiency indicates what fraction of the energy of the XUV
radiation absorbed in the atmosphere goes into heating. The heating
efficiency has been assigned values from 0 to 1 in various studies.
Calculations of the photoelectron kinetics are necessary for a full
determination of this quantity. Since photoelectrons are suprathermal
particles and have a non-Maxwellian velocity distribution, the Boltzmann
equation must be solved in order to carry out such calculations. We did
this in~\cite{Shematovich-2014,Ionov-2015} using a model in which the
electron kinetics were determined using Monte Carlo simulations. This
model was used to compute the height profiles of the heating intensity
and heating efficiency in the atmosphere of the hot Jupiter HD~209458b
based on the density height profiles of~\cite{Yelle-2004}. The mean
heating efficiency over all heights in the atmosphere was 0.14. However,
this computed profile of the heating efficiency is valid only when the
components of the atmosphere are distributed in accordance with the
results of~\cite{Yelle-2004}, and other component distributions could
give rise to different heating-efficiency profiles. Therefore, correct
computations of the dynamics of the atmosphere of a hot Jupiter require
the development of a complex, self-consistent model that includes
computation of the dynamics of suprathermal particles, chemical
reactions, and gas dynamics. The construction of such a model was the main
task of our current study.

\section{Model}

We present here a self-consistent, aeronomic model for the upper
atmosphere of a hot Jupiter with taking into account the suprathermal
photoelectrons. This numerical model includes a Monte Carlo module,
chemical module, and gas-dynamical module. The structure of the model is
shown schematically in Fig. 1. In the Monte Carlo module, the intensity
of heating in the atmosphere and the rates of ionization, dissociation,
and excitation of the atmospheric components are computed based on the
initial distributions of the concentrations of the atmospheric components
and the temperature. The concentrations of the atmospheric components are
computed in the chemical module, based on the reaction rates computed in
the Monte Carlo module. In the gas-dynamical module, variations of the
macroscopic characteristics of the atmosphere --- density, velocities,
temperature --- are computed using the heating rate.

\begin{figure*} 
\begin{center}
\includegraphics[width=110 mm]{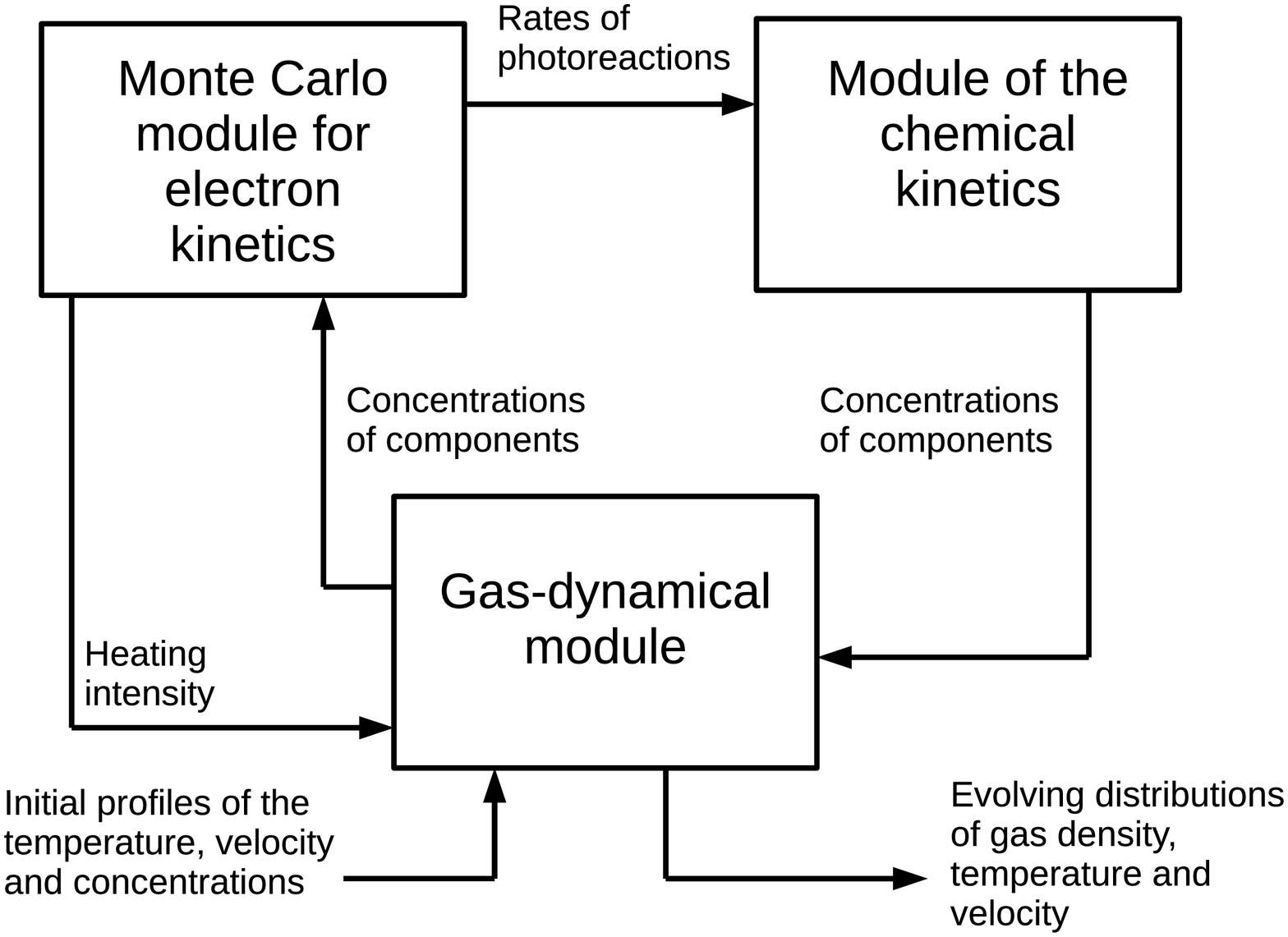}
\caption{Chart of the model.}
\end{center}
\label{fig1}
\end{figure*}

The transport and kinetics of photoelectrons in an exoplanetary upper
atmosphere dominated by hydrogen and helium were computed using the Monte
Carlo model of~\cite{Shematovich-2008,Shematovich-2010}, adapted for a
hydrogen atmosphere. The model includes the reactions~\eqref{eq_ioniz}
and the transport of suprathermal electrons in the atmosphere. The energy
of an electron formed during a collision and subsequent ionization is
chosen in accordance with the procedure described
in~\cite{Garvey-Green-1976,Jackman-1977,Garvey-1977}. Accordingly, the
kinetics and transport of photoelectrons are described using the
Boltzmann equation,
\begin{equation}
\vec{v} \frac{\partial }{\partial \vec{r}} f_e + \vec{s} \frac{\partial
}{\partial \vec{v}} f_e = Q_{e,photo}(\vec{v}) + Q_{e,secondary} (\vec{v}) +
\sum_{M=H_2,H,He} J(f_e,f_M),
\end{equation}
where $f_e(r,v)$  and $f_M(r,v)$ are the distribution functions for the
velocities of the electrons and components of the ambient atmospheric
gas, respectively. The transport of electrons in the force field
$\vec{s}$ of the planet is described on the left-hand side of the
equation. The term $Q_{e,photo}$ on the right-hand side of the kinetics
equation describes the rate of formation of fresh electrons via
photoionization, and the term $Q_{e,secondary}$ the formation of
secondary electrons via ionization by photoelectrons. The collision
integrals for elastic and inelastic interactions between the electrons
and the ambient atmospheric gas $J(f_e,f_M)$ are written in the standard
form, assuming that the atmospheric gas is characterized by a local
equilibrium Maxwellian velocity distribution.

A detailed description of the realization of the Monte Carlo model for
the transport of photoelectrons in the planetary atmosphere is presented
in~\cite{Marov-1996,Shematovich-2008,Shematovich-2010}. We note only
that this realization used experimental and computational data for the
cross sections and scattering-angle distributions for elastic, inelastic,
and ionization collisions between electrons and H$_2$, He, and H,
selected from the sources listed in~\cite{Shematovich-2010}. The partial
and total rates of ionization by the flux of photoelectrons were
specified using standard formulas based on computed distribution
functions for the electrons in the thermosphere.

The rate at which the energy of radiation and photoelectrons is
transformed into internal energy in each of the photoreactions and in
reactions with secondary electrons is computed in the module. The energy
of the suprathermal photoelectrons that is transformed into thermal
energy is calculated separately. Thus, the modeling results can be used
to determine the heating function of the atmosphere.

A system of chemical-kinetics equations is solved in the chemical module.
This reaction network includes 19 reactions involving nine components: H,
H$_2$, e$^-$,  H$^+$, H$_2^+$,  H$_3^+$, He, He$^+$, HeH$^+$. The
reaction-rate constants were taken from~\cite{Munoz-2007}. Since the
resulting system of differential equations is stiff, it was solved using
the DVODE program package.  The main channel for radiative cooling is
emission by the ion H$_3^+$. The dependence of the intensity of this
emission on the temperature was taken from~\cite{Neale-1996}, and the
cooling function was calculated on this basis.

The basis of the gas-dynamical module is the numerical code described
in~\cite{Pavlyuchenkov-2015}. This code was first created to compute the
collapse of a protostellar cloud, and was adapted by us to be suitable
for a planetary atmosphere before we applied it. A one dimensional
spherically symmetrical adiabatic system of gas-dynamical equations is
solved in the model using the equation of state of an ideal gas:
\begin{eqnarray}
&&\dfrac{1}{\rho}=r^2\dfrac{\partial r}{\partial q} \nonumber \\
&&\dfrac{dr}{dt}  = v \nonumber \\
&&p=(\gamma-1)\rho \varepsilon \\
&&\dfrac{dv}{dt}  = - r^2\dfrac{\partial p}{\partial q} - G\dfrac{M}{r^2} 
\nonumber  \\
&&\dfrac{d\varepsilon}{dt} = -p\dfrac{\partial }{\partial q}\left( r^2 v\right) 
\nonumber
\end{eqnarray}
where $q$ is the Lagrangian mass coordinate, which is related to a mass
element in a spherical layer $\Delta m$ by the expression $\Delta m =
4\pi\Delta q$, $r$ is the radial coordinate, $t$ the time, $\rho$ the
density, $v$ the velocity, $p$ the pressure, $\varepsilon$ the specific
thermal energy, $\gamma$ the adiabatic index, and $M$ the mass of the
planet. The system of equations was solved using an implicit, fully
conservative difference scheme described in~\cite{Samarsky-1992}. The
computations were carried out on a Lagrangian grid, that is, with movable
cell boundaries.

The adiabatic index was taken to be $\gamma = 5/3$ in the computations.
This value is valid for a monoatomic gas; the real adiabatic index will
vary with height, since the atmosphere consists partially of molecules of
hydrogen, as well as the ions H$_2^+$, H$_3^+$ and HeH$^+$. To test
whether these variations significantly influenced the computation
results, we carried out a test computation with a variable adiabatic
index. This test computation showed that deviations of the adiabatic
index from 5/3 did not appreciably influence the resulting profiles.

Since the density in the atmosphere falls off exponentially with radius,
the use of a grid that is uniform in the mass variable means that cells
near the surface of the planet are much smaller than cells in the upper
part of the computed region. This leads to a reduction in the time step
determined from the Courant condition. Thus, the uniform grid leads to a
low computation rate and low spatial resolution in the upper part of the
atmosphere, where key processes are occurring. To solve this problem, the
numerical method used in~\cite{Pavlyuchenkov-2015} was modified for
computations on a non-uniform grid. In our computations, we used a grid
that was non-uniform in the mass coordinate: the mass of the cells falls
off with height in accordance with a geometric progression. The index of
this progress was chosen empirically to be 0.986. Artificial viscosity
was introduced into the scheme in order to suppress oscillations arising
in dense layers of the atmosphere. The value of this viscosity was chosen
to be the minimum sufficient to suppress non-physical oscillations.

At each time step, after implementing the gas-dynamical module, the
thermal energy and pressure values were renewed taking into account the
heating function $\Gamma$ and cooling function $\Lambda$ obtained in the
chemical module and Monte Carlo module:
\begin{equation}
\dfrac{d\varepsilon}{dt} = \Gamma-\Lambda, \label{main6}
\end{equation}
This approach is widely used and is known as splitting according to the
physical processes.

The planet HD~209458b was chosen as the object of study. This is the
first detected transiting hot Jupiter, for which the results of several
observations and numerous data modeling its atmosphere are available.

\begin{figure*}
\centering
 \includegraphics[width=80 mm]{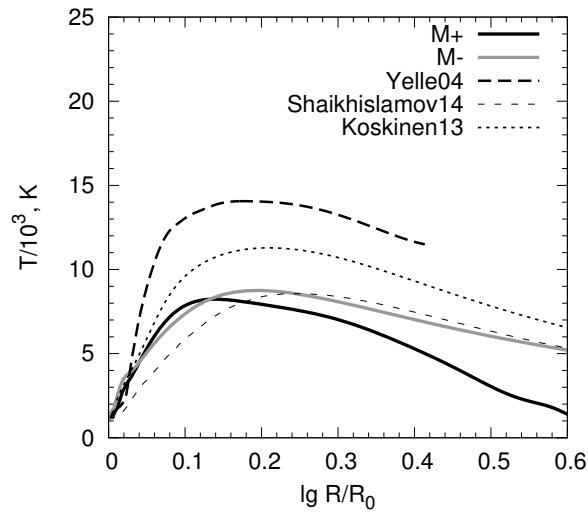}
\caption{Height profiles of the
temperature for the models M+, M-, Koskinen13, Shaikhislamov14, and
Yelle04. Data for this and subsequent figures
can be downloaded from \url{https://github.com/Pantarktik/ARepIonovEtAl2017}.}
\label{fig_t}
\end{figure*}

\begin{figure*} 
\centering
\includegraphics[width=80 mm]{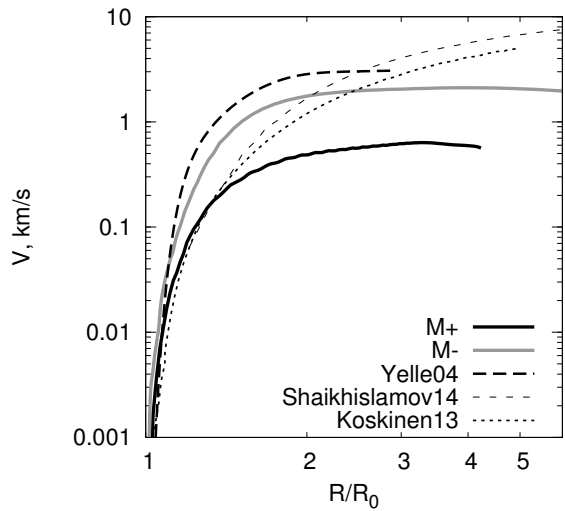}
\caption{Height profiles of the velocity for models M+, M-,
Koskinen13, Shaikhislamov14, and Yelle04.}
\label{fig_v}
\end{figure*}

We chose the following initial conditions for the computations. The
atmosphere was taken to be isothermal at a temperature of 1300~K, which
corresponds to the equilibrium temperature at a distance from the star
equal to the semi-major axis of the orbit of HD~209458b. The density in
the atmosphere was distributed according to a barometric law, and the gas
velocity was initially equal to zero. The lower boundary of the
computational domain was rigidly fixed at a distance equal to one
planetary radius, with the conditions for reflection implemented. The
upper boundary was not fixed; i.e. the atmosphere could both expand and
contract during the simulations. The outer boundary condition was
provided by the external pressure, which was taken to be equal to the gas
pressure of the stellar wind at the corresponding orbital radius. Taking
the stellar-wind parameters presented in~\cite{Withbroe-1988}, the
external pressure is equal to $p_{ex}=1.6 \times 10^{-6}$ dyne/cm$^2$.
The atmosphere initially consisted of molecular hydrogen and helium, with
a particle number fraction of helium equal to 0.15. One of the input
parameters of the model is the total mass of the atmosphere. This was
chosen empirically, such that the part of the atmosphere heated by XUV
radiation fell into the computational domain. In our case, the mass of
the atmosphere was $1 \cdot 10^{18}$~g. The number density at the lower
boundary was initially $7 \times 10^{11}$~cm$^{-3}$. The spectrum of the
star in the UV was taken to be the same as the spectrum of the current
Sun, taken from~\cite{Huebner-1992} and recalculated for a distance of
0.045 AU. The computations were carried out until a steady state was
reached.

Further, we compare the results for the model taking into account
reactions involving suprathermal electrons (M+) and the model without
these reactions (M-), that is, with an enhanced heating intensity. We
also used the results obtained~\cite{Yelle-2004} (Yelle04), 
\cite{Koskinen-2013} (Koskinen13), and~\cite{Shaikhislamov-2014}
(Shaikhislamov14) in this comparison.

\section{Results}

\begin{figure*}
\centering
 \includegraphics[width=80 mm]{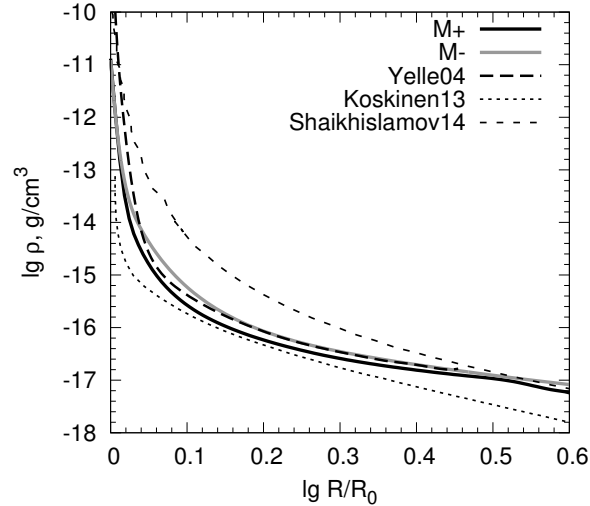}
\caption{Height profiles of the density for models M+, M-,
Koskinen13, Shaikhislamov14, and Yelle04.}
\label{fig_rho}
\end{figure*}

Figures~\ref{fig_t}--\ref{fig_rho} show the computed height profiles of
the temperature, velocity, and density for all the models considered.
Since including suprathermal particles in the computations leads to a
decrease in the heating intensity, we expect that including such
particles should result in an appreciable decrease in the temperature of
the atmosphere, as well as a reduction of the gas velocity, since the gas
velocity is increased by heating.

We can see that our computations gave results that are qualitatively
similar to the results of other studies. The height profiles of the
temperature for models M+ and M-, as well as for models from other
studies, are shown in Fig.~\ref{fig_t}. In agreement with expectations,
taking into account photoelectrons leads to a reduction in the
temperature of the atmosphere, with the difference between our two models
growing with increasing radius. While the maximum temperature in model M-
is roughly 9000~K, including suprathermal particles in the model reduces
this,  maximum temperature to 6000~K. Among the three models with which
we compared our own computations, the best agreement is observed for the
model Skaihislamov14. The temperatures in the Yelle04 and Koskinen13
models are several thousand Kelvin higher. However, the maximum
temperature is located at roughly the same distance from the center in
all the models: 1.3--1.5~$R_0$.

The velocity profiles shown in Fig.~\ref{fig_v} are also qualitatively
similar to the profiles obtained in other studies, and encompass the same
range of velocities. Only at high heights do the models Koskinen13 and
Skaihislamov14 display velocities exceeding the results of our model M-.
The difference between models M+ and M- is especially large for the
velocity: the gas velocity in the model without photoelectrons is a
factor of a few higher.

Fig.~\ref{fig_rho} compares the density profiles for models M+, M-,
Koskinen13, Shaikhislamov14, and Yelle04. Taking into account
photoelectrons does not strongly influence the density profile, and the
curves for models M+, M-, and Yelle04 virtually coincide. 

The mass loss rate of the atmosphere can be estimated from the density
and velocity at a given radius using the formula
\begin{equation}
 \dot{M} = 4 \pi \rho(R) \upsilon(R) R^2
\end{equation}

The computed mass loss rate does not change at heights above 1.2
planetary radii. The mass loss rate in model M-
is~$\dot{M} \approx  4 \cdot 10^{10}$~g/s. The mass loss rate
at the same heights in model M+ is~$ \dot{M} \approx 8 \cdot 10^9$~g/s.
The mass loss rate in model M- agrees with the results for the model
Shaikhislamov14 ($7 \cdot 10^{10}$~g/s) and essentially coincides with
the model Koskinen13 ($4 \cdot 10^{10}$~g/s). Taking into account the
photoelectrons leads to a reduction in the mass loss rate by a factor of
a few.

As was shown in~\cite{Bisikalo-2013a,Bisikalo-2013b,Cherenkov-2014}, the
regime and mass loss rate of the atmosphere are determined not only by
the state of the atmosphere, but also by the parameters of the stellar
wind. Therefore, the derived parameters of the atmosphere can be used as
boundary conditions for three-dimensional gas-dynamical computations
modeling the interaction between the planetary atmosphere and the stellar
wind.

\section{Conclusion} 

We have carried out simulations of the upper atmosphere of the exoplanet
HD~209458b both including and excluding reactions involving suprathermal
electrons. Including suprathermal particles leads to a strong reduction
in the gas velocity. The mass loss rate falls by a factor of five in the
computations with photoelectrons. Thus, we have demonstrated that
suprathermal particles are an important factor in exoplanetary
atmospheric processes, and neglecting them will lead to substantial
errors in estimates of the parameters of the atmosphere.

\section{Acknowledgements}

This work was supported by the Russian Science Foundation (D.I. by
Project 14-12-01048 and V.S. by Project 15-12-30038) and a grant of the
President of the Russian Federation for the Support of Scientific Schools
of the Russian Federation (project NSh-9576.2016.2).

\bibliographystyle{mn2e} 
\bibliography{stherm2}

\end{document}